\documentclass[3p,times,procedia]{elsarticle}
\usepackage{nupha_ecrc}
\biboptions{square, comma, sort&compress, numbers}
\usepackage{subfigure}

\volume{00}

\firstpage{1}

\journalname{Nuclear Physics A}

\runauth{}

\jid{nupha}

\jnltitlelogo{Nuclear Physics A}

\usepackage{amssymb}

\usepackage[figuresright]{rotating}

\begin{document}

\begin{frontmatter}



\dochead{XXVIIIth International Conference on Ultrarelativistic Nucleus-Nucleus Collisions\\ (Quark Matter 2019)}

\title{Interplaying mechanisms behind inclusive jet $R_{AA}$ and extraction of jet energy loss distributions}


\author{Y. He$^1$\corref{ref1}}
\author{S. Cao$^2$, W. Chen$^1$, T. Luo$^1$, L.-G. Pang$^{3,4}$ and X.-N. Wang$^{1,3,4}$}
\cortext[ref1]{heyayun@mails.ccnu.edu.cn}

\address{$^1$Key Laboratory of Quark and Lepton Physics (MOE) and Institute of Particle Physics, Central China Normal University, Wuhan 430079, China \\
$^2$Department of Physics and Astronomy, Wayne State University, Detroit, Michigan 48201, USA\\
$^3$Physics Department, University of California, Berkeley, California 94720, USA\\
$^4$Nuclear Science Division Mailstop 70R0319,  Lawrence Berkeley National Laboratory, Berkeley, CA 94740,USA}

\begin{abstract}
\label{abstract}
The observed inclusive jet suppression in heavy-ion collisions at LHC has a very weak $p_{T}$ dependence over a large range of $p_{T}$ = 50-1000 GeV and is almost independent of the colliding energy, though the initial energy density of the bulk medium has increased from $\sqrt{s}$ = 2.76 to 5.02 TeV by about 20\%. This interesting phenomenon is investigated in the linear Boltzmann transport (LBT) model for jet propagation in an event-by-event 3+1D hydro background. We show that the $p_{T}$ dependence of jet $R_{AA}$ is determined by the initial spectrum in $p+p$ collisions and $ p_{T} $ dependence of jet energy loss. Furthermore, jet energy loss distributions for inclusive jet and $ \gamma-$jet at both LHC energies are extracted directly from experimental data through the state-of-art Bayesian analysis. The averaged jet energy loss has a weak $p_{T}$ dependence and the scaled jet energy loss distributions have a large width, both of which are consistent with the LBT simulations and indicate that jet quenching is caused by only a few out-of-cone jet medium scatterings.
\end{abstract}


\end{frontmatter}

\section{Introduction}
\label{sec:intro}
Jet quenching is a powerful tool to investigate the properties of quark-gluon plasma (QGP) created in high-energy nucleus-nucleus collisions \cite{qgp1, qgp2, qgp3}. A larger suppression for inclusive jet at $ \sqrt{s} = 5.02 $ TeV is expected than at $ \sqrt{s} = 2.76 $ TeV since the bulk hadron density at the mid-rapidity increases by about $20 \%$ \cite{hd1, hd2}. However, one observes a weak $p_{T}$ and colliding energy dependence of jet $ R_{AA} $ at the large $p_{T}$ range in Pb+Pb collisions \cite{RAA:ATLAS2760, RAA:ATLAS5020, RAA:CMS2760}. The interesting behavior of jet $R_{AA}$ is caused mainly by the $ p_{T} $ dependence of jet energy loss due to jet quenching in addition to the spectrum of initial jet production. Similar behavior of heavy flavor $R_{AA}$ is also predicted at LHC energies ~\cite{heavy_RAA}. Furthermore, despite the fact that the extraction of jet transport coefficient is important to understand QGP properties \cite{qhat}, jet energy loss distributions should provide us with extra information about jet medium interaction. It is therefore important for us to extract jet energy loss distributions in heavy-ion collisions. 

\section{The LBT model}
\label{sec:lbt}
The Linear Boltzmann Transport (LBT) model \cite{LBT1, LBT2, LBT3, LBT4, Luo:2018PLB, He:2019PRC, He:2019PRL} is based on the Boltzmann transport equation including both elastic scattering according to the leading order perturbative Chromodynamics (pQCD) and the inelastic scattering from a high twist approach \cite{inelastic1, inelastic2}. The elastic and inelastic probability, which are implemented together to ensure unitarity, are assumed to obey the Poisson distribution in each time step $\Delta  t = 0.1 fm/c$. The effective strong coupling constant $\alpha_{S}$ in the LBT model is fixed for strong interaction regulated by the Debye screening mass, and set to 0.15 by fitting to experimental data, 

The initial jet shower partons are generated within PYTHIA 8. They will propagate according to the Boltzmann transport equation in the medium, whose evolution is simulated by the (3+1)D CLVisc hydrodynamic model with fully event-by-event fluctuation initial conditions \cite{hydro1, hydro2}. Jet shower partons, jet-induced medium recoil and radiated gluons are tracked and allowed to rescatter. Jet-induced medium back-reaction which is denoted as "negative" particles is also included and will be subtracted during final jet reconstruction to make sure the global energy-momentum conservation for each scattering. The LBT model has a linear approximation that only the interaction with medium thermal partons is taken into consideration, and it is only valid if the density of the medium excitation is much smaller than the background density. To relax such an approximation, a coupled LBT and hydrodynamic (CoLBT-hydro) model has been developed \cite{LBThydro}, in which the deposited energy and momentum of the jet shower partons is considered as a source term of the hydrodynamic evolution in real time.

\section{Weak $p_{T}$ and $\sqrt{s}$ dependence of the inclusive jet suppression}
\label{sec:raa}
\begin{figure}[t]
    \centering
    \subfigure[]{\includegraphics[width=4.9cm]{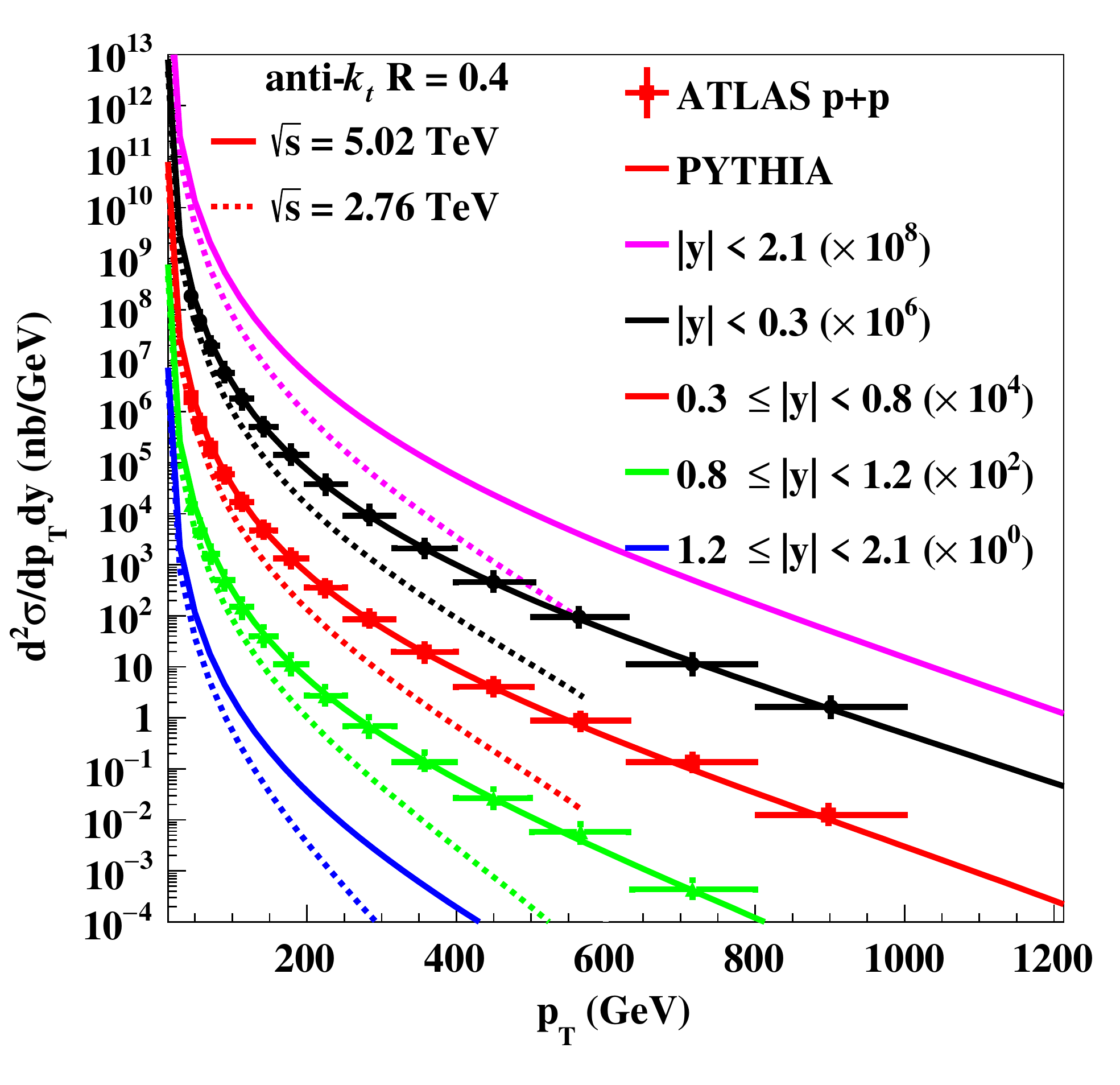}}
    \subfigure[]{\includegraphics[width=4.9cm]{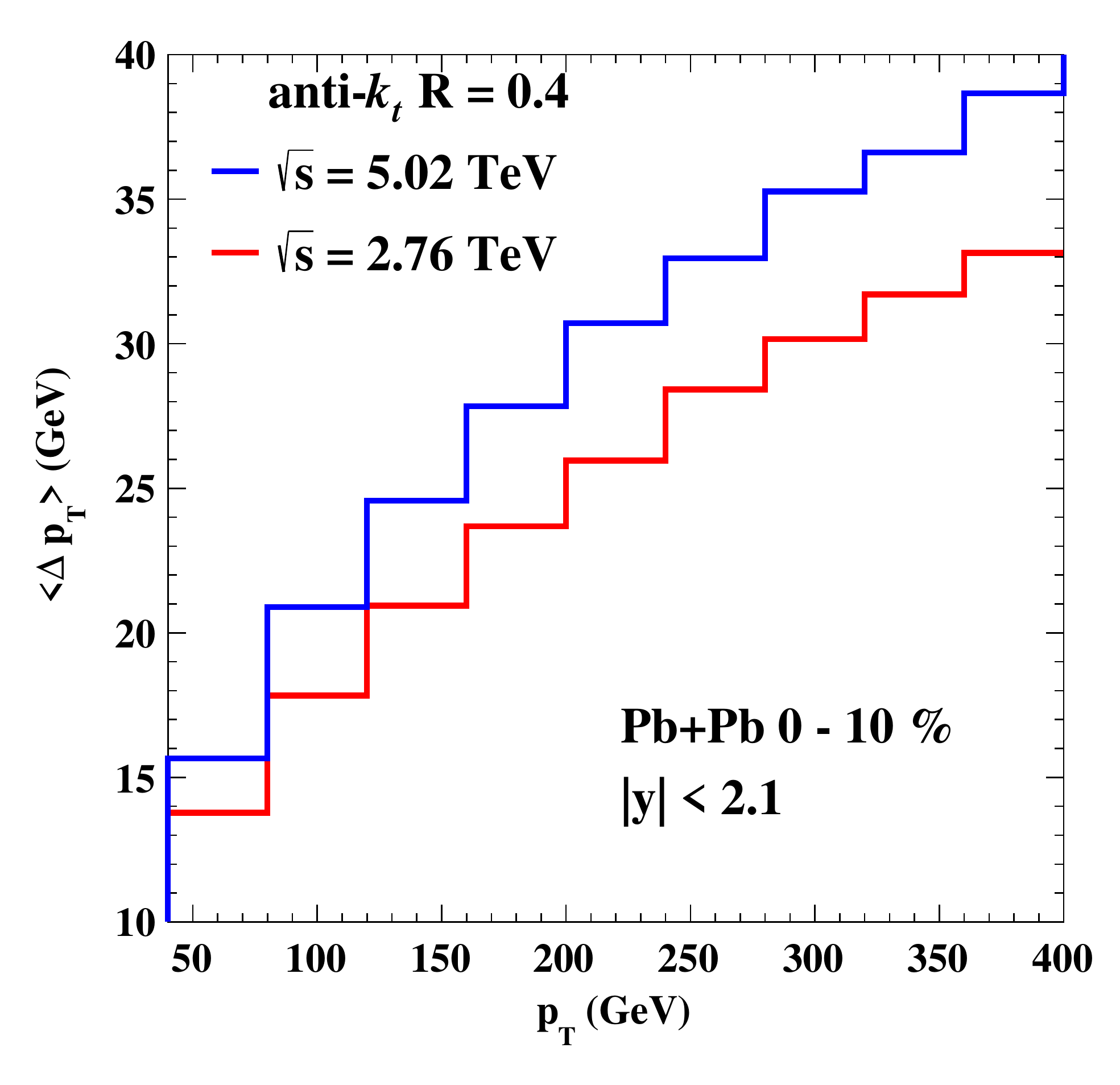}}
    \subfigure[]{\includegraphics[width=4.9cm]{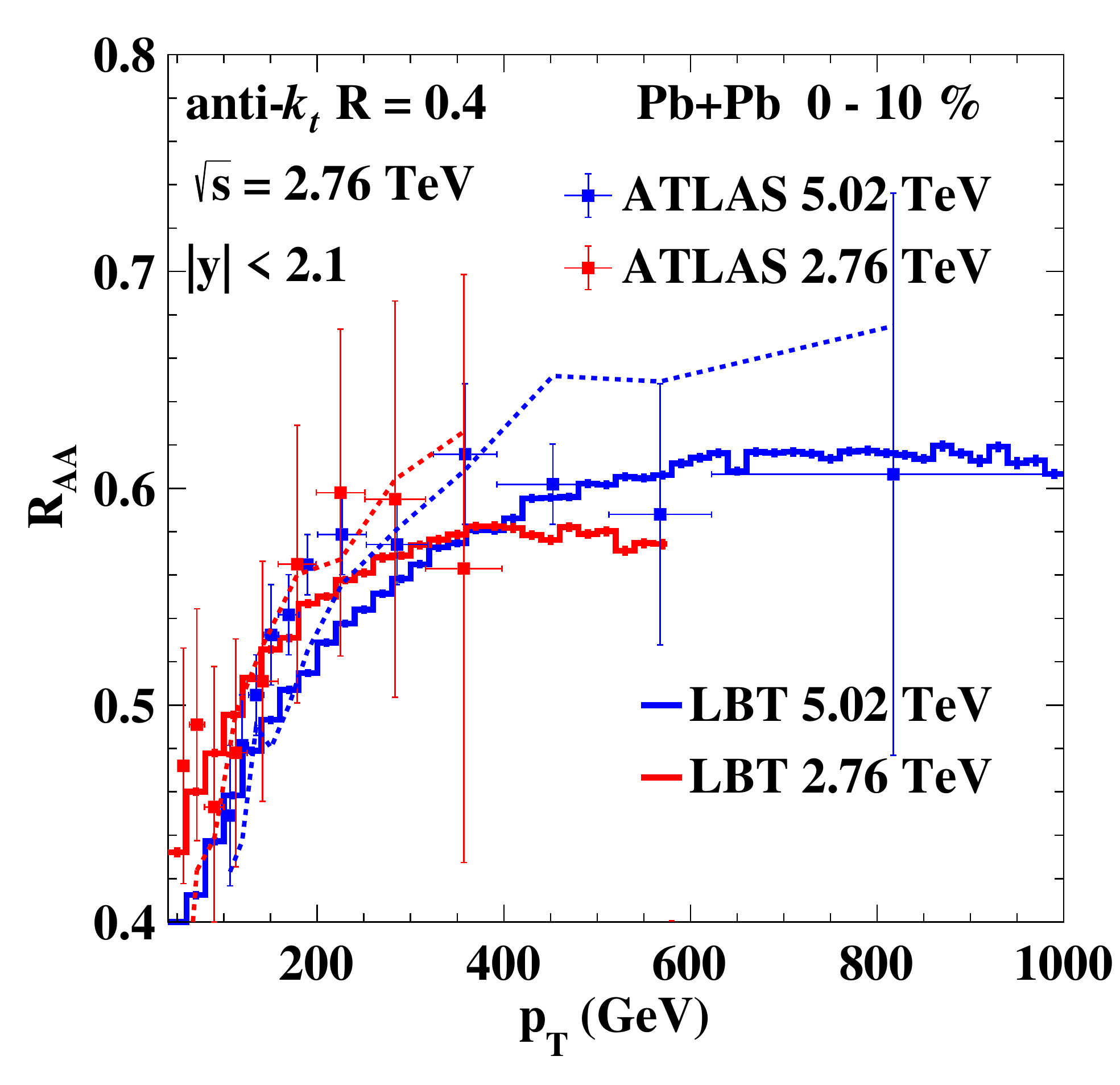}}
    \vspace{-0.3cm}
    \caption{(Color online) (a) the inclusive jet double differential cross section at different rapidity bins with jet cone size $R=0.4$ at $\sqrt{s} = 2.76$ TeV (dashed) and $\sqrt{s} = 5.02$ TeV (solid) at $p+p$ collisions from PYTHIA 8 (lines) compared to experimental data. Results for different rapidity bins are scaled by successive power of $10^{2}$. (b) $p_{T}$ dependence of the averaged jet $p_{T}$ loss and (c) the inclusive jet suppression factor $R_{AA}$ with jet radius $R = 0.4$ at $\sqrt{s} = 5.02$ TeV (blue) and $\sqrt{s} = 2.76$ TeV (red) at Pb+Pb  collisions at $0 - 10 \%$ centrality in jet rapidity $|y| < 2.1$. The dashed lines are suppression factors obtained by shifting the $p+p$ spectrum by the averaged $p_{T}$ loss according to Eq. (\ref{shiftRAA}).}
    \label{fig:raa}
\end{figure}

The shower partons in the LBT model are generated from $p+p$ collisions in the vacuum on the approximation that the parton distribution functions are not modified by the nuclear medium in the high energy limit. Fig. \ref{fig:raa} (a) shows the double differential jet cross section with jet cone size $R = 0.4$ at both $\sqrt{s} = 2.76$ TeV and $\sqrt{s}  = 5.02$ TeV in p+p collisions. One can see that PYTHIA 8 can well describe the experimental data \cite{RAA:ATLAS2760, RAA:ATLAS5020}, and that the spectra at $\sqrt{s} = 5.02$ TeV are much flatter than at $\sqrt{s} = 2.76$ TeV, due to the difference of parton distribution functions at different colliding energies.

Given the reference in p+p collisions, we can calculate the inclusive jet suppression factor~\cite{He:2019PRC}. Fig. \ref{fig:raa} (b) shows the averaged jet $p_{T}$ loss as a function of the initial jet $ p_{T} $ at $\sqrt{s} = 5.02$ TeV and $\sqrt{s} = 2.76$ TeV. There is indeed an $18 \%$ enhancement of the averaged jet energy loss at the higher colliding energy. However, Fig. \ref{fig:raa} (c) shows that the jet suppression factors at both colliding energies are about the same and have a weak $p_{T}$ dependence at the high $p_{T}$ range in both experimental data and the LBT calculations. The reason is that the influence from the initial jet spectrum in $p+p$ collisions competes with that from jet quenching. The initial spectrum is much flatter as jet $p_{T}$ increases at $\sqrt{s} = 5.02$ TeV than at $\sqrt{s} = 2.76$ TeV. This flatness of jet $p_{T}$ spectrum cancels the effect of large jet energy loss at $\sqrt{s} = 5.02$ TeV and gives almost the same values of jet $R_{AA}$ at both colliding energies. In the same way, the combination of $p_{T}$ dependence of initial jet spectrum and jet energy loss result in a weak $ p_{T}$ dependence of jet $ R_{AA} $ in the large $ p_{T} $ range.

One can estimate jet $R_{AA}$ according to Eq. (\ref{shiftRAA}) when $ \langle \Delta p_{T} \rangle / p_{T} $ is small ,
\begin{equation}
    R_{AA} (p_{T}) \approx  \frac{d\sigma^{jet}_{p+p} (p_{T} + \langle \Delta p_{T} \rangle )}{d\sigma^{jet}_{p+p} (p_{T})}.
    \label{shiftRAA}
\end{equation}
Here jet $R_{AA}$ at a given jet $p_{T}$ can be approximately expressed as the ratio of the initial cross section at that $p_{T}$ shifted by the averaged jet $p_{T}$ loss over that without a shift, shown as the dashed lines in Fig. \ref{fig:raa} (c), which  gives a consistent description of the experimental data within the uncertainty.

\section{Bayesian extraction of jet energy loss distributions}
\label{sec:bayes}
Fig.~\ref{fig:raa} (c) shows that jet $R_{AA}$ calculated according to shifting the initial jet cross section by an averaged jet energy loss still has a large deviation from the LBT results without such an approximation. This indicates the fluctuation of jet energy loss should be significant and that jet energy loss distribution should be taken into account instead. Indeed, jet nuclear modification factor $R_{AA}$ can be expressed as the convolution of the flavor-averaged (quark and gluon) jet energy loss distribution $W_{AA}$ and the initial jet production cross section in $p+p$ collisions in the pQCD factorization framework~\cite{He:2019PRL}
\begin{equation}
  R_{AA} (p_{T}, R)= \frac{1}{d\sigma^{pp}_{jet}(p_{T}, R)} \int d\Delta p_T W_{AA} (\Delta p_T, p_T + \Delta p_T, R) \, d\sigma^{pp}_{jet} (p_T + \Delta p_T, R).
  \label{eq:RAA}
\end{equation}
Motivated by the LBT results~\cite{He:2019PRC, LBT3}, we parametrize $W_{AA}$ as a normalized $\Gamma$ function as a function of a scaled variable $x=\Delta p_{T} / \left \langle \Delta p_{T} \right \rangle$, and the averaged jet energy loss as a log function modulated by a power law of $ p_{T}$,
\begin{equation}
  W_{AA} (x) = \frac{\alpha^{\alpha} x^{\alpha - 1} e^{-\alpha x}}{\Gamma(\alpha)}, \quad \quad \quad \quad
  \langle \Delta p_{T}  \rangle ( p_{T} ) = \beta p_{T}^{\gamma} \log(p_{T}).
  \label{eq:WAA}
\end{equation}

\begin{figure}[h]
    \centering
    \subfigure[]{\includegraphics[width=7.0cm]{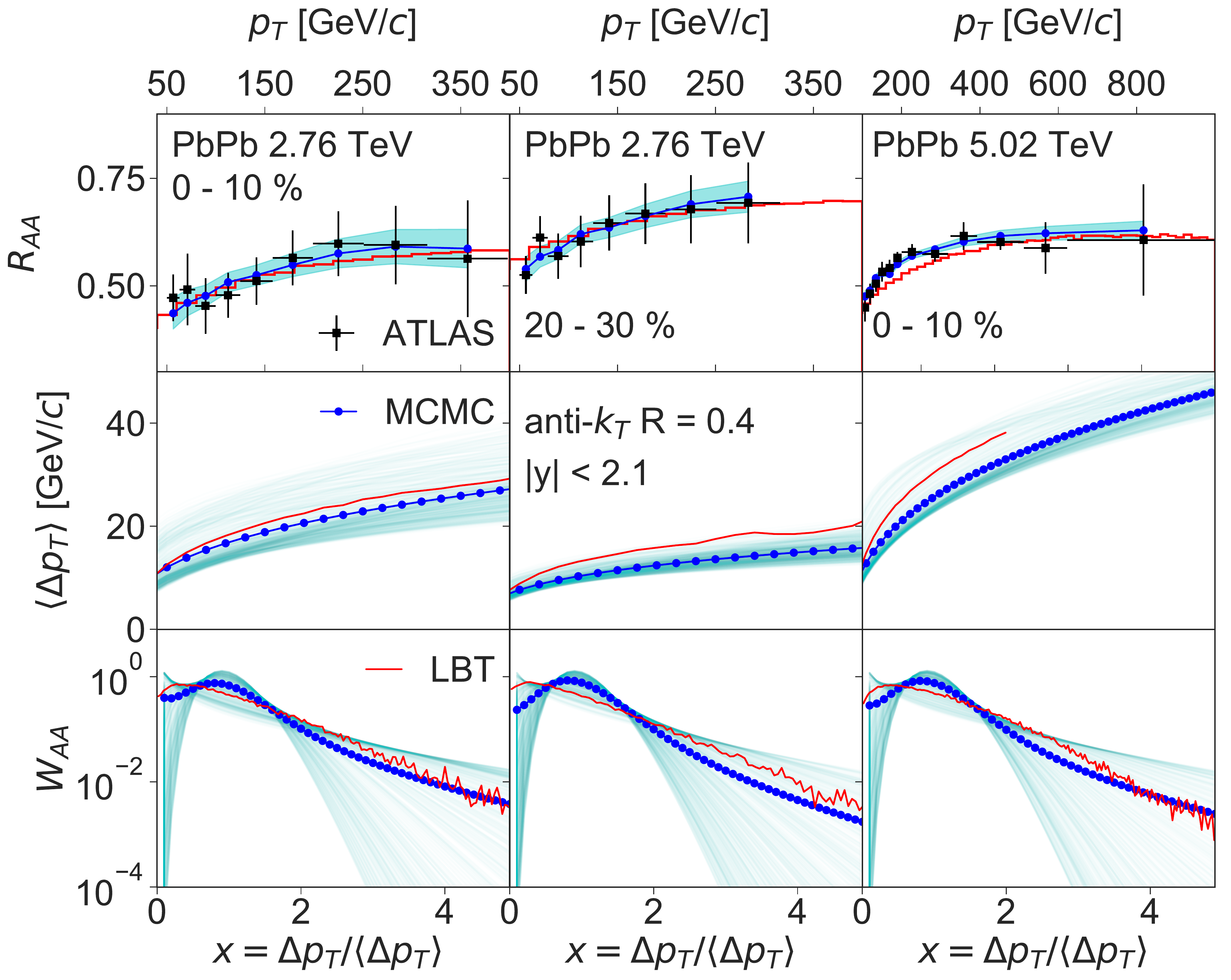}}
    \subfigure[]{\includegraphics[width=7.0cm]{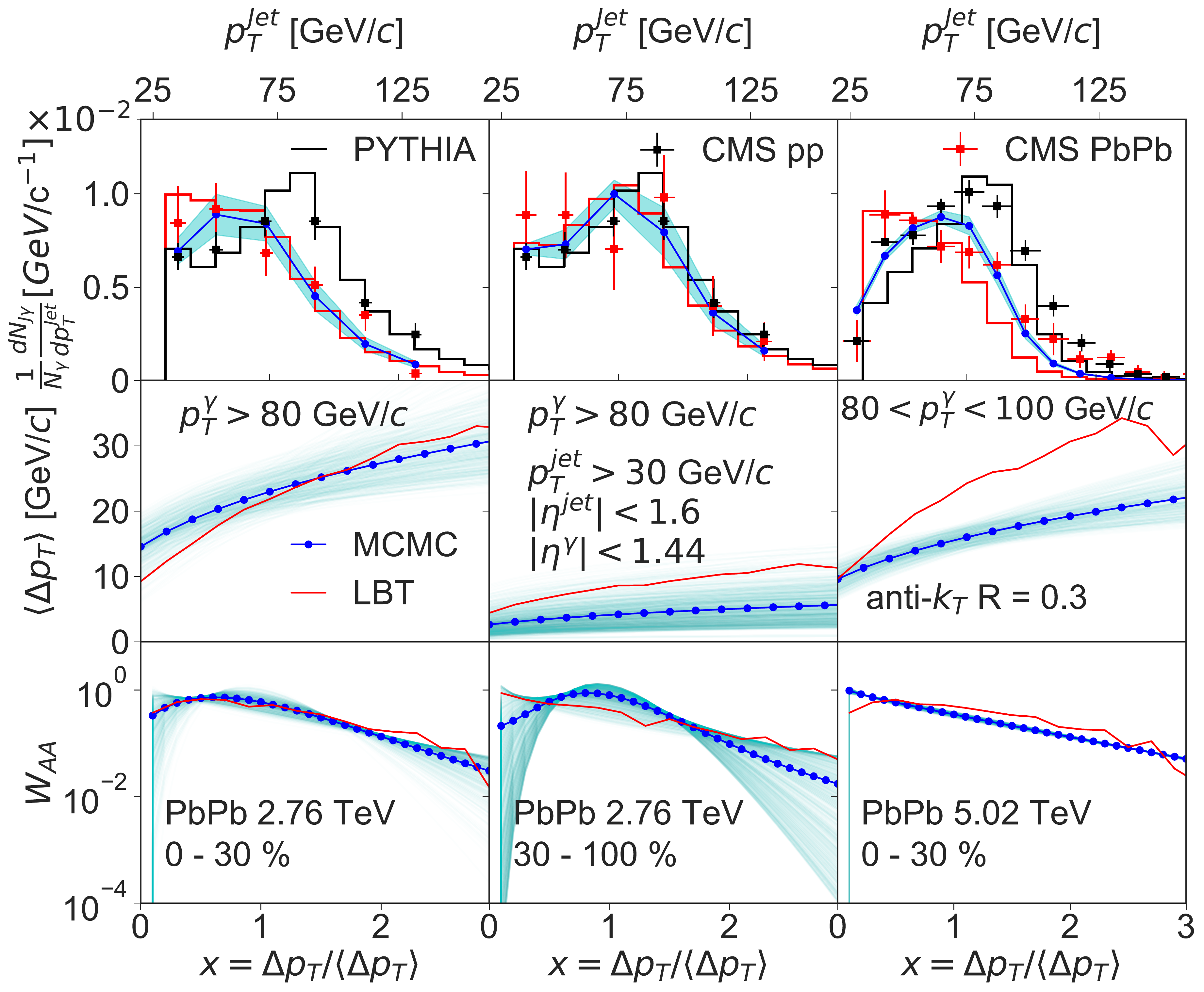}}
    \vspace{-0.3cm}
    \caption{(Color online) (a) inclusive jet $ R_{AA} $, (b) $ \gamma- $jet distributions, averaged jet energy loss and scaled jet energy loss distributions from Bayesian analysis (shadow lines) and the LBT simulations (red lines) compared to data at different colliding energies at different centrality bins. }
    \label{fig:mcmc}
\end{figure}

\begin{figure}[h]
    \centering
    \subfigure[]{\includegraphics[width=7.0cm]{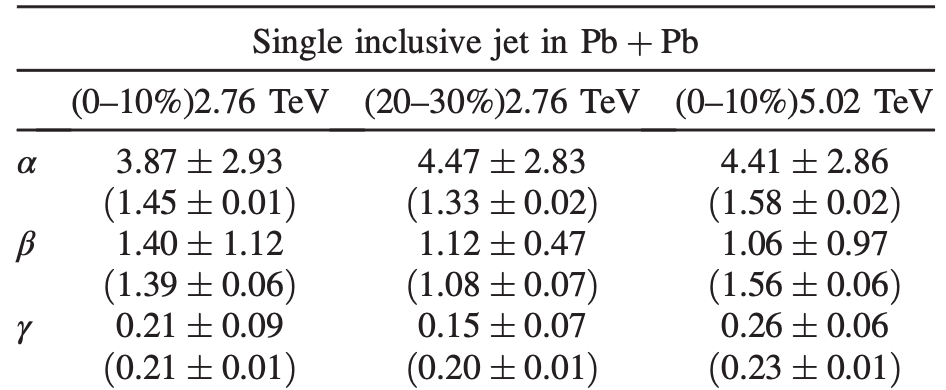}}
    \subfigure[]{\includegraphics[width=7.0cm]{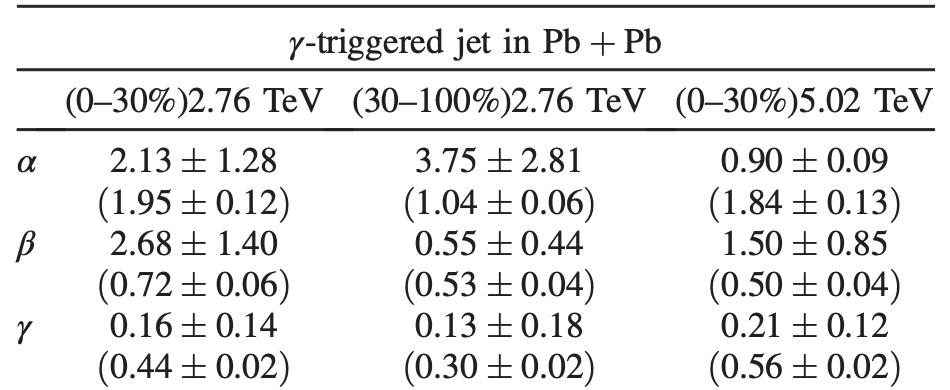}}
    \vspace{-0.3cm}
    \caption{(Color online) Table of the mean and standard deviation of the parameters extracted from experimental data and from the LBT model (in parentheses) for inclusive jet (a) and $ \gamma-$jet (b) at different colliding energies at different centrality bins. } 
    \label{fig:table}
\end{figure}

In the LBT model, jet $R_{AA}$ and $W_{AA}$ are calculated by solving the Boltzmann equation. On the other hand, one should be able to extract $W_{AA}$ directly from the experimental data on jet cross section in both $p+p$ and $A+A$ collisions by employing the state-of-art Bayesian analysis. Fig.~\ref{fig:mcmc} shows inclusive jet $ R_{AA} $ ($ \gamma- $jet distributions), averaged jet energy loss and scaled jet energy loss distributions from Bayesian analysis and the LBT simulations compared to data at different colliding energies in different centrality bins. We can see Bayesian analysis can even better describe the data than the LBT calculations. In Fig.~\ref{fig:table} list the mean and standard deviation of the parameters extracted from experimental data and from the LBT model. They are consistent within the uncertainties. The small value of $ \alpha$ can be interpreted that there are only a few out-of-cone jet medium scatterings along jet propagation path.

\section{Conclusion}
\label{sec:conc}
The inclusive jet suppression factor is not only dependent on jet quenching, but also significantly influenced by the initial jet production spectra. By performing Bayesian analysis of the experimental data, the extracted jet energy loss distributions indicate there are only a few out-of-cone jet medium interactions in Pb+Pb collisions at LHC.

\section{Acknowledgement}
This work is supported by the NSFC under grant Nos. 11935007, 11890714, 11221504, NSF within the JETSCAPE Collaboration and under grant No. ACI-1550228, and ACI-1550300, and U.S. DOE under Contract Nos. DE-AC0205CH11231 and DE-SC0013460. Computations are performed at the DOE NERSC.


\begin{thebibliography}{00}
\bibitem{qgp1}
    STAR Collaboration, \emph{Nucl. Phys. A} {\bf 757}, 102 (2005).

\bibitem{qgp2}
    PHENIX Collaboration, \emph{Nucl. Phys. A} {\bf 757}, 184 (2005).

\bibitem{qgp3}
    B. Muller, J. Schukraft, and B. Wyslouch, \emph{Annu. Rev. Nucl. Part. Sci.} {\bf 62}, 361 (2012).


\bibitem{hd1}
    ALICE Collaboration, \emph{Phys. Lett. B} {\bf 726}, 610 (2013).

\bibitem{hd2}
    ALICE Collaboration, \emph{Phys. Lett. B} {\bf 772}, 567 (2017).


\bibitem{RAA:ATLAS2760}
    ATLAS Collaboration, \emph{Phys. Rev. Lett.} {\bf 114}, 072302 (2015).

\bibitem{RAA:ATLAS5020}
ATLAS Collaboration, \emph{Phys.Lett. B} {\bf 790} 108 (2019).

\bibitem{RAA:CMS2760}
    CMS Collaboration, \emph{Phys. Rev. C} {\bf 96}, 015202 (2017).

\bibitem{heavy_RAA}
M. Djordjevic and M. Djordjevic, Phys. Rev. C \textbf{92}, 024918 (2015).


\bibitem{qhat}
    JET Collaboration, \emph{Phys. Rev. C} {\bf 90}, no. 1, 014909(2014).

\bibitem{LBT1}
    H. Li, F. Liu, G. l. Ma, X. N. Wang and Y. Zhu, \emph{Phys. Rev. Lett.} {\bf 106}, 012301 (2011).

\bibitem{LBT2}
    X. N. Wang and Y. Zhu, \emph{Phys. Rev. Lett.} {\bf 111}, no. 6, 062301 (2013).

\bibitem{LBT3}
    Y. He, T. Luo, X. N. Wang and Y. Zhu, \emph{Phys. Rev. C} {\bf 91}, 054908 (2015) Erratum:[\emph{Phys. Rev. C} {\bf 97}, no. 1, 019902 (2018)]. 

\bibitem{LBT4}
    S. Cao, T. Luo, G. Y. Qin and X. N. Wang, \emph{Phys. Lett. B} {\bf 777}, 255 (2018).

\bibitem{Luo:2018PLB}
T. Luo, S. Cao, Y. He, and X. N. Wang, \emph{Phys. Lett. B} {\bf 782}, 707 (2018).

\bibitem{He:2019PRC}
    Y. He, S. Cao, W. Chen, T. Luo,  L. G. Pang and X. N. Wang, \emph{Phys. Rev. C} {\bf 99}, 054911 (2019). 


\bibitem{He:2019PRL}
Y. He, L. G. Pang and X. N. Wang, \emph{Phys. Rev. Lett.} {\bf 122} 252302 (2019).


\bibitem{inelastic1}
    X. F. Guo and X. N. Wang, \emph{Phys. Rev. Lett.} {\bf 85},3591 (2000).

\bibitem{inelastic2}
    X. N. Wang and X. F. Guo, \emph{Nucl. Phys. A} {\bf 696}, 788 (2001).

\bibitem{hydro1}
    L. G. Pang, Q. Wang and X. N. Wang, \emph{Phys. Rev. C} {\bf 86}, 024911 (2012).

\bibitem{hydro2}
    L. G. Pang, H. Petersen and X. N. Wang, \emph{Phys. Rev. C} {\bf 97}, no. 6, 064918 (2018).

\bibitem{LBThydro}
    W. Chen, S. Cao, T. Luo, L. G. Pang and X. N. Wang, \emph{Phys. Lett. B} {\bf 777}, 86 (2018).



\end{thebibliography}
\end{document}